\documentclass[twoside,12pt]{article}
\usepackage{a4}
\usepackage{amsmath,amsfonts,latexsym}

\newtheorem{satz}{Theorem}[section]
\newtheorem{defi}[satz]{Definition}

\newtheorem{bem}[satz]{Remark}
\newtheorem{lemma}[satz]{Lemma}
\newtheorem{koro}[satz]{Corollary}
\newtheorem{bsp}[satz]{Example}
\newtheorem{assumption}[satz]{Assumption}
\newtheorem{obdef}[satz]{Observation/Definition}
\newtheorem{conclusion}[satz]{Conclusion}
\newtheorem{ob}[satz]{Observation}
\newtheorem{consequences}[satz]{Consequences}

\newcommand{\mcal}{\mathcal}

\newcommand{\tit}{\textit}
\newcommand{\tbf}{\textbf}
\newcommand{\C}{\mathbb{C}}
\newcommand{\N}{\mathbb{N}}
\newcommand{\R}{\mathbb{R}}
\newcommand{\m}{M_{\rho}(\R^m)}
\newcommand{\W}{W_2(\R^m)}
\newcommand{\lp}{l^p(\C),1\leq p\leq \infty}
\newcommand{\bew}{ \bf Proof\rm: {}}
\newcommand{\bewende}{$ \hfill \Box $\vspace{\baselineskip}\par}

\begin{document}
\allowdisplaybreaks
\hfuzz=0.3cm

\thispagestyle{empty}
\begin{center}
\vspace*{1.0cm}

{\LARGE{\bf Perturbation Theory of Schr\"odinger Operators in\\
    Infinitely Many Coupling Parameters}}

\vskip 1.5cm

{\large {\bf Manfred
Requardt\footnote{email:requardt@Theorie.Physik.Uni-Goettingen.de}\quad
Anja Schl\"omerkemper\footnote{current adress: Max-Planck-Instiute for
Mathematics in the Sciences, Inselstr. 22-26, 04103 Leipzig,
Germany, email: Anja.Schloemerkemper@mis.mpg.de} }} 

\vskip 0.5 cm 

Institut f\"ur Theoretische Physik \\ 
Universit\"at G\"ottingen \\ 
Bunsenstrasse 9 \\ 
37073 G\"ottingen \quad Germany

\end{center}

\vspace{1cm}

\begin{abstract}
  In this paper we study the behavior of Hamilton operators and their
  spectra which depend on infinitely many coupling parameters or,
  more generally, parameters taking values in some Banach
  space. One of the physical models which motivate this framework is a
  quantum particle moving in a more or less disordered medium. One
  may however also envisage other scenarios where operators are
  allowed to depend on interaction terms in a manner we are going to
  discuss below. The central idea is to vary the occurring infinitely
  many perturbing potentials independently.
  As a side aspect this then leads naturally to the analysis of a couple of
  interesting  questions of a more or less purely mathematical flavor
  which belong to the field of infinite dimensional holomorphy or
  holomorphy in Banach spaces. In this general setting we study in
  particular the stability of selfadjointness of the operators under
  discussion and the analyticity of eigenvalues under the condition
  that the perturbing potentials belong to certain classes.

\end{abstract} \newpage \setcounter{page}{1}

\section{Introduction}
The physical aim of the paper is the investigation of properties of
Hamilton operators which depend on infinitely many coupling
parameters, $\beta_i$, or perturbing potentials, $V_i$, i.e. we want
to study Hamilton operators of the form:
\begin{equation}H(\beta)=H_0+\sum_{i=1}^{\infty}\beta_iV_i
\end{equation}
with some unperturbed Hamiltonian $H_0$, which properties are
frequently assumed to be known. In typical cases $H_0$ is some free
Hamiltonian like e.g. the Laplacian $-\Delta$ or some relatively
well-behaved standard Hamiltonian of the form
\begin{equation}H_0=-\Delta+U \end{equation}
with a fixed interaction potential $U$.

In the general situation $\beta:=(\beta_i)$ varies in a certain
infinite dimensional sequence space like e.g. $l^p$ with $1\le
p\le\infty$, where $l^{\infty}$, i.e. the sequences which do not
necessarily decay at infinity, is particularly interesting on
physical grounds as we are primarily interested in perturbations which
extend with same strength to spacial infinity.

The physical motivation to study this class of model Hamiltonians is
the following: We envisage a quantum mechanical particle moving in an
infinitely extended background consisting, say, of a more or less
disordered array of atomic potentials. It is then an interesting
scenario, both from the physical and mathematical point of view, to
test the response of the particle to independent variations of the
coupling strengths or potentials making up the array. A typical case
in point is a particle moving in a regular crystal which is then
deformed or develops more or less irregularly distributed
defects. Furthermore, possible applications to disordered media in
general are obvious.

Among the various aspects one can or should investigate are several of
a more mathematical or fundamental flavor as e.g. \tit{selfadjointness
  questions} or \tit{analyticity properties of eigenvalues or the
  spectrum in general}. These are the problems we will mainly adress
in the following in order to set the stage as we are presently not aware
whether such questions have been dealt with in this generality in the
past. Let us note in this context that up to now the main thrust of
investigations has rather gone into the study of \tit{random
  Hamiltonians}, an approach which is in some sense complementary to
the one we will develop in this paper.

In adressing problems of this general kind one soon realizes that
intricate mathematical questions do emerge which do not belong to the
standard arsenal of mathematical physics as e.g. \tit{infinite
 dimensional holomorphy} or \tit{holomorphy on Banach
 spaces}. Furthermore one has to deal with Taylor series having a
countable infinity of independent variables. In other words, this is
yet another example how a natural physical problem quickly leads into
some very advanced fields of pure mathematics.  

This suggests the following organization of the paper: As a warm-up
exercise we will treat in the next section the special case of bounded
perturbations within the context of general operator theory and
develop a couple of useful mathematical tools and concepts.  This
abstract approach, that is, making only very general assumptions on
the class of potentials under discussion, leads, perhaps surprisingly,
to certain technical problems if one attempts to apply it in a next
step to unbounded operators. These problems are briefly discussed at the
beginning of section 3.

In section 3 we then show that a seemingly appropriate concrete class
of potentials is the so-called \tit{Stummel class} if one is willing
to adopt a more concrete setting, i.e. working within a concrete
Hilbert space of functions and studying concrete Hamilton operators.
We then return to more abstract considerations and give a brief review
of \tit{infinite dimensional holomorphy} or \tit{complex analysis in
  Banach spaces}. This allows to treat Hamiltonians, which depend on
infinitely many independent coupling constants (or more generally,
coupling parameters belonging to some Banach space), and in particular
their \tit{perturbation theory} in a more systematic way in section 5.

\section{Concepts and Tools}

The first main step consists in providing criteria so that the
Hamiltonian $H=H_0+\sum\beta_iV_i$ is again selfadjoint given the
selfadjointness of $H_0$. To begin with, this problem shall be studied
with the help of a simple class of perturbations for which the
well known additional technical intricacies of the more general
situation are expected to be absent.

\begin{assumption} Let $H_0$ be a selfadjoint operator on the
Hilbert space $\mcal{H}$. The $V_i,\: i\in \mathbb{N}$, are
assumed to be {\em uniformly bounded}, i.e.
\begin{equation}\|V_i\|\le v<\infty\end{equation}
for all $i$ and some $v\in\R$.    
\end{assumption}

The problem is to guarantee that the infinite sum over the potentials
is again a well-defined operator and, in this example, a
bounded operator. In general it may easily happen that such an
infinite sum is no longer defined on certain vectors in the Hilbert
space (e.g. if the potentials tend to cluster to much around certain
points in coordinate space). In order to prevent this one has to take
some precautions. A sufficient condition which furthermore
has a clear geometric or physical meaning is the following: 

\begin{defi}[Finite Intersection Property] \label{fip}
Let the $V_i$'s be
linear operators on $\mcal{H}$. We say they have the {\em finite
  intersection property} if the following holds:\\
There exists a projection valued probability measure $P$ on
$\mcal{B}(\mathbb{R}^m)$, the Borel $\sigma$-algebra over
$\mathbb{R}^m$, so that:
\begin{enumerate}
\item For each $V_i$ exists a Borel set $\Omega_i$ with
  $P_{\Omega_i}V_i=V_i$.
\item For each $i\in\mathbb{N}$ let $I_i$ be the index set
  $\{i\neq j\in\mathbb{N}:\Omega_i\cap\Omega_j\neq\emptyset\}$. Then
 $\#(I_i)$ is uniformly bounded in $i$ by some $n_0<\infty$.
\end{enumerate}
\end{defi}

Remark: Note that in the general case the $V_i$ can be almost
arbitrary localized operators acting in some abstract Hilbert space. 
In the same sense the correlation of the projectors with certain sets in some
space $\mathbb{R}^m$ can, while of course being physically motivated,
be fairly indirect from a mathematical point of view.

The above entails that to each given $V_i$ there exist at most $n_0$
projectors $P_j$ (short hand for $P_{\Omega_j}$) so that $P_jV_i\neq
0$ since $P_jP_i=0$ if $\Omega_j\cap\Omega_i=\emptyset$. By the same
argument there exist at most $n_0$ potentials $V_i$ so that $P_jV_i\neq
0$.

\begin{bsp}[Multiplication Operators] If $V_i$ are
multiplication operators on $L^2(\mathbb{R}^m)$ they have the finite
intersection property if
\begin{enumerate}
\item the support of $V_i$ is contained in a Borel set $\Omega_i$ and
\item if for these sets condition 2 of Definition 2.2 holds with the
  $P_i$'s being indicator functions.
\end{enumerate}
\end{bsp}

Remark: The above condition entails that the potentials are
sufficiently scattered in coordinate space. This can also be enforced
by slightly different conditions like e.g. the following. Assume that,
given an arbitrary $x\in\mathbb{R}^m$ and a ball around $x$ with some
fixed diameter, only uniformly finitely many potentials meet this ball
when $x$ varies over $\mathbb{R}^m$. We will come back to this variant
in section 3.
\vspace{0.5cm}

We now proceed as follows. With the help of the polar decomposition of
$V_i$ we have
\begin{equation}\|V_i\psi\|=\|U_i|V_i|\psi\|=\||V_i|\psi\|
\end{equation}
with $|V_i|:=(V_i^*V_i)^{1/2}$. Applying the corresponding
projector $P_i$ to a selfadjoint $V_i$ we get
\begin{equation}P_iV_i=V_i=V_i^*=V_iP_i \end{equation}
hence
\begin{equation}\||V_i|\psi\|^2\leq v^2\cdot\|P_i\psi\|^2
\end{equation}
and
\begin{equation}|V_i|\leq v\cdot P_i \end{equation}
since for positive operators $A^2\leq B^2$ implies $A\leq B$ (meaning
$(\psi|A\psi)\leq(\psi|B\psi)$). For finitely many $V_i$ then
follows that
\begin{equation}\sum_1^n |V_i|\leq v\sum_1^n P_i. \end{equation}
\vspace{0.5cm}

Given the sequence of potentials $V_i$ or projectors $P_i$ we now make
a disjoint refinement $\{\Omega_j^{\prime}\}$ of the class of sets
$\{\Omega_i\}$. This then yields a corresponding refinement of the
class of projectors which are now orthogonal by construction
(projection valued measure, hence $\Omega_i^{\prime}\cap\Omega_j^{\prime}=\emptyset$
implies $P_i^{\prime}\cdot P_j^{\prime}=0$). The construction is accomplished in the
following way. First of all we can restrict ourselves to an arbitrary
but fixed candidate of the class $\{\Omega_i\}$ which we for
convenience call $\Omega_0$. We then make the following
definition.

\begin{obdef} For each given $x\in\Omega_0$ there
exists a unique maximal index set $I_x\subset \{0,1,\ldots,k\}$ with
$\Omega_1,\ldots,\Omega_k$ being the sets intersecting the start set $\Omega_0$
and $x\in\Omega_j$ for $j\in I_x$. We call the subset of elements of $\Omega_0$
having the same maximal index set $I$, $\Omega_I$. This construction
yields a (disjoint) partition
into a finite number of Borel sets of the start set $\Omega_0$. In the
same way we can proceed with the other
sets of the class $\{\Omega_i\}$, thus arriving at the disjoint
partition $\{\Omega_j^{\prime}\}$.
\end{obdef}

Remark: Note that the sets of the refinement basically consist of
certain intersections and corresponding complements within the class
of the initial sets.

\begin{lemma}: The sets $\Omega_j^{\prime}$ represent a disjoint
partition of the original class with a given $\Omega_i$ being
resolved into at most $2^{n_0}$ disjoint sets where $n_0$ was the
upper bound on the number of sets $\Omega_j$ intersecting a given
$\Omega_i$. By the same token we get a resolution into mutually
orthogonal projectors with
\begin{equation}P_i=\sum_1^{k_i}P_j^{\prime},\quad k_i\leq
  2^{n_0}.\end{equation}
\end{lemma}

\bew By assumption at most $n_0$ sets can intersect a given
$\Omega_i$. Furthermore one of the disjoint sets can result from the
maximal index set $\{0\}$, corresponding to the complement in
$\Omega_0$ of the union of all the intersections of $\Omega_i$ with
$\Omega_0$. The optimal scenario can then be estimated by counting the
number of subsets of a $(n_0)$-set (i.e. having $n_0$ elements), which
is $2^{n_0}$. To this we have to add $1$ for the above complement and
subtract $1$ for the empty set counted in $2^{n_0}$. This proves the
above estimate.  \mbox{}\bewende

For a finite sum of $P_i$'s we then have
\begin{equation}\sum_1^n P_i\leq n_0\cdot\sum_j^k
  P^{\prime}_j\end{equation}
with $k$ being a number between $n$ and $2^{n_0}\cdot n$. Note that each
$P_j^{\prime}$ can occur at most $n_0$ times on the rhs. By construction
$\sum_1^k P_j^{\prime}$ is again a projector (in contrast to the lhs),
hence has norm one and we get (cf. equation (8)): 
\begin{equation}\sum_1^n |V_i|\leq v\cdot\sum_1^n P_i\leq v\cdot
  n_0\cdot\boldsymbol{1}\end{equation}
implying
\begin{equation}\|\sum_1^n|V_i|\|\leq v\|\sum_1^n P_i\|\leq v\cdot
  n_0.\end{equation}
With $|V_i|$ positive the sum on the lhs is monotonely increasing with
a global norm bound given by the rhs. As for selfadjoint $V_i$ 
 $|(\psi|V_i\psi)|\leq (\psi||V_i|\psi)$ holds, we get
\begin{equation}|(\psi|\sum_1^n V_i\psi)|\leq(\psi|\sum_1^n
  |V_i|\psi)\end{equation}
and
\begin{equation}(\psi|\sum_1^n|V_i|\psi)\leq\|\psi\|\cdot\|\sum_1^n|V_i|\psi\|\leq
  v\cdot n_0\cdot\|\psi\|^2.\end{equation}
Thus the sequence $\sum_1^n V_i$ converges weakly to a bounded
operator $V=\sum_1^{\infty} V_i$ because the lhs is a Cauchy sequence.

\begin{conclusion}\mbox{}
\begin{enumerate}
\item Under the assumptions made above $\sum_1^{\infty}V_i$ can be
  defined as a weak limit and is again a selfadjoint bounded operator
\item This implies by standard reasoning (see e.g. \cite{RS1} or
  \cite{Ka}) that $H=H_0+\sum V_i$ is again a selfadjoint operator on
  the domain of $H_0$ where here and in the sequel, unless otherwise
  noted, unspecified summation always means summation from $1$ to $\infty$.
\end{enumerate}
\end{conclusion}

\begin{koro} This applies in particular to a potential
\begin{equation}V=\sum\beta_iV_i,\quad\beta_i\in\mathbb{R},\quad\beta:=(\beta_i)\in
  l^{\infty}(\mathbb{R}), \quad V_i \;\text{as above}.\end{equation}
\end{koro}

\begin{bem}\mbox{}
\begin{enumerate}
\item If the $V_i$ are real multiplication operators on
some $L^2(\mathbb{R}^m)$ the above sum can be shown to converge even
in the strong sense. The underlying abstract reason for this stronger
property lies in the fact that now all the $V_i$ do automatically
commute which allows for certain technical manipulations of sums which
do not seem to be possible in the more general case (the proof can be
found in \cite{Schloem}).
\item Furthermore, in this concrete case there exists a more
  straightforward variant on the above proof. The operator bound of
  $V_i$ is the $\operatorname{ess\,sup} |v_i(x)|$. By assumption at most
  $n_0$ $v_i$ meet at a point $x$. The operator norm of $\sum V_i$ is
  hence bounded by $v\cdot n_0$.
\end{enumerate}
\end{bem}
\tbf{Warning}: In the generic case the above convergence is not in
operator norm. Assuming e.g. that $\|V_i\|\geq\varepsilon> 0$ for all
$i$, the sequence of sums $\sum_1^n V_i$ is evidently not a Cauchy
sequence in norm.
\vspace{0.5cm}

In the typical physical situation the occurring potentials are
frequently not of (strictly) finite range but decay at infinity with a
certain rate. The \tit{finite intersection property} introduced above
emulates to some extent such a \tit{finite range condition}. It is
therefore an interesting question to what extent an infinite range of
the potentials under discussion can be admitted. We make the following
assumption.

\begin{assumption}[Infinite Range] \label{Infinite Range}
We assume that each potential
$V_i$ can be decomposed as
\begin{equation}V_i=V_i^a+V_i^b\end{equation}
with $V_i^a$ fulfilling the finite intersection property. We assume
further that with the help of the methods developed in this paper
\begin{equation}H^{\prime}=H_0+\sum V_i^a\end{equation}
can be given a rigorous meaning as a selfadjoint operator.
We want to
impose a condition on $V_i^b$ so that 
\begin{equation}\|\sum V_i^b\|<\infty\end{equation}
i.e., so that
\begin{equation}H=H_0+\sum V_i^a+\sum V_i^b\end{equation}
is a well defined selfadjoint operator with $\sum V_i^b$ being a bounded
perturbation of $H^{\prime}$. On the other side the $V_i^b$ need not
fulfill the finite intersection property. We assume that the $V_i$ are
multiplication operators in $L^2(\mathbb{R}^m)$ with the $V_i^b$
centered around points $R_i$ and decaying in the following way:
\begin{equation}|V_i^b(x)|\leq const/(1+|R_i-x|)^k\end{equation}
for some $k>0$ 
and with the $R_i$ distributed in $\mathbb{R}^m$ according to
\begin{equation}|R_i-R_j|>2A>0\end{equation}
if $i\neq j$.
\end{assumption}
Our strategy is to show that under these conditions
\begin{equation}\sum |V_i^b(x)|\leq\sum const/(1+|R_i-x|)^k\leq
  B<\infty\end{equation}
for all $k\geq k_0$. To this end we prove the following simple
lemma:
\begin{lemma}
With $x\in\mathbb{R}^m$ and $K_R(x)$ a ball of
radius $R$ centered at $x$ there are at most $(R+A)^m/A^m$ points $R_i$
in $K_R(x)$ if $|R_i-R_j|>2A$. Correspondingly one can estimate the
number of points in a spherical shell $K_{[R,R+d)}$ around $x$ of radii
$R,R+d$ with $R>A$. We have
\begin{equation}\text{\#}(R_i)\leq[(R+d+A)^m-(R-A)^m]/A^m.\end{equation}
\end{lemma}

\bew Let $\{R_i\}$ be the set of points lying in $K_R(x)$. Draw a
sphere of radius $A$ around each $R_i$. The corresponding balls do not
intersect and we can hence estimate:
\begin{equation}\text{\#}(R_i)\cdot c_m\cdot A^m\leq
  c_m\cdot(R+A)^m,\end{equation}
where $c_m$ is the volume of the unit sphere in $\mathbb{R}^m$. From
this we can conclude
\begin{equation}\text{\#}(R_i)\leq (R+A)^m/A^m.\end{equation}
In the same way we prove the second statement.
\bewende

We can now proceed as follows:
\begin{equation}\sum |V_i^b(x)|=\sum_{R_i\in
    K_l}|V_i^b(x)|+\sum_{n\geq l}\sum_{R_i\in 
   K_{[n,n+1)}}|V_i^b(x)|\end{equation}
where $n,l\in \mathbb{N}$ with $l$ arbitrary but fixed so that
$l>A$. The rhs can be estimated so that
\begin{eqnarray}
  \sum |V_i^b(x)| &\leq& C_0+\sum_{n\geq l}(const/(1+n)^k)\cdot
  [(n+1+A)^m-(n-A)^m]/A^m\nonumber\\
  &\leq& C_0^{\prime}+C_1\sum_n
  n^{(m-1-k)}
\end{eqnarray}
with $C_0,C_0^{\prime},C_1$ being constants independent of the point
$x$ (note that the leading $n^m$-powers vanish. Furthermore we have absorbed
sums over terms with a smaller power than $m-1-k$ in the
constants). This sequence is convergent for $k>m$, hence:

\begin{ob} For potentials fulfilling the criteria of
assumption \ref{Infinite Range} the sum over infinite range
potentials, $\sum V_i^b$, 
yields a bounded operator if 
\begin{equation}|V_i^b(x)|\leq const/(1+|R_i-x|)^k,\quad
  k>m\end{equation}
with $m$ being the space dimension.
\end{ob}

\section{The Stummel Class}

It is tempting to try to proceed in the same abstract way as developed
in section 2 by simply admitting more general classes of potentials or
operators $V_i$. Our original idea  was to employ the famous criterion
of \tit{Kato smallness} in its abstract form  in order to arrive at
selfadjoint perturbations of a given selfadjoint start Hamiltonian
(see e.g. \cite{Ka},\cite{RS2}).
\begin{defi}
Let $H_0$, $V$ be operators on $\mcal{H}$. $V$ is
called $H_0$-bounded with relative bound $a$ if
\begin{enumerate}
\item $D(V)\subset D(H_0)$
\item $\|V\psi\|\leq
    a\|H_0\psi\|+b\|\psi\|$
\end{enumerate}

with $a,b$ real and $a$ understood as the infimum of all these
constants.
\end{defi}

Remark: A corresponding condition can be formulated in the \tit{weak}
(i.\ e.\ \tit{form}) sense.
\begin{satz} 
With $H_0,V$ as above $H=H_0+V$ is a closed or
selfadjoint operator on $D(H_0)$ if $H_0$ is closed or selfadjoint and
$V$ is symmetric in the latter case provided that $a<1$.
\end{satz}

It would now be natural to assume the $V_i$ to be \tit{Kato-small} in
the above sense and then try to show the same for $\sum_{i=1}^\infty V_i$. But to
our surprise, irrespectively of the direction of attack, an approach
along these abstract lines was not yet successful due to technical
intricacies in the manipulation and interchange of (infinite) sums and
norm estimates. As a consequence we choose, for the time being, a more
concrete approach in this section and consider a certain (in fact
large) class of admissible potentials on some
$L^2(\mathbb{R}^m)$. 
\begin{defi}[Stummel class] With $v$ a measurable on
$\mathbb{R}^m$ with respect to standard Lebesgue measure we define for
each $\rho\in\mathbb{R}$
 \begin{equation} M_{v,\rho}(x) = \left\{ \begin{array}{l@{\quad:\quad}l}
                             \big(\int_{|x-y|\leq 1} |v(y)|^2 |x-y|^{\rho -m}\, d^m y\big)^\frac{1}{2} 
                             & \rho<m, \\ 
                             \big(\int_{|x-y|\leq 1} |v(y)|^2 \,d^m y\big)^\frac{1}{2} & \rho\geq m.
                           \end{array} \right. \end{equation}
The corresponding Stummel class is given by
 \begin{equation} \m := \{ v:\R^m \rightarrow \C \;: \sup_{x \in \R^m}
 M_{v,\rho}(x):=M_{v,\rho}<\infty \}. \end{equation}
\end{defi}

This class was introduced by Stummel in \cite{St}. A textbook
treatment can e.g. be found in \cite{Weid}. Its properties has also
been exploited in various papers of B.\ Simon (see e.g.\ \cite{Si}).

\begin{lemma}[without proof] $\m$ is a vector space and
$M_{\rho_1}\subset M_{\rho_2}$ for $\rho_1\leq\rho_2$.
\end{lemma}

\begin{bsp} \label{Ex}
\begin{equation}L^2(\R^m)+L^{\infty}(\R^m)\subset \m,\quad
  L^2_{loc}(\R^m)\subset \m
\end{equation}
for $\rho\geq m$.
\end{bsp}

As to the reason for the choice of this particular class we would like
to make some comments. Typically mathematical
physicists are accustomed to atomic potentials which consist of a 
\tit{singular part}, having
a few singularities of a certain degree away from infinity and perhaps
a certain decaying tail extending to infinity. The lhs of Example \ref{Ex}
is a typical case in point. These are the classes for which a lot of
estimates can be found in the literature (see e.g. \cite{RS2}) and
which lead to a whole bunch of selfadjointness criteria. The
potentials we want to discuss however are of a more intricate
type. In our scenario the singularities generically extend to infinity
as the particle is assumed to move in an infinitely extended
(disordered) medium. As far as we can see, most of the standard
estimates do apply only to the above mentioned simpler class of atomic
potentials (at least without modifications). On the other side, as can
be seen from Definition 3.1, the Stummel condition is essentially a
local estimate, that is, it is relatively insensitive to the number of
singularities and their position in space. Therefore it seems to be
more suitable for our purposes at the moment.

We remarked already in section 2 that there exist variants on the
\tit{finite intersection property} given in Definition \ref{fip} which
may turn out to be more suitable in specific contexts. This is
e.g. the case for the Stummel class.

\begin{defi}[Variant on Finite Intersection Property] \label{Variant}
Given $x\in\R^m$ and a ball around $x$ with radius one, there are only
uniformly finitely many potentials $V_i$ (with respect to each $x$)
which meet this ball. The bound being denoted by $n_1$.
\end{defi}

Remark: For \tit{well behaved sets} $\Omega_i$ or supports of $V_i$ all
these conditions are essentially equivalent. On the other side, there
may be extreme situations where the one or the other turns out to be
better adapted.

\begin{satz} With $v_i(x)$ in $\m$ for all $i$ so that
$M_{v_i,\rho}<\infty$ uniformly in $i$ and $\{v_i\}$ fulfilling the
intersection property in the sense of Definition \ref{Variant}
\begin{equation}\sum_1^{\infty} \beta_iv_i\in \m\end{equation}
holds for $\beta\in l^p(\C),\; 1\leq p\leq\infty$.
\end{satz}

\bew By assumption there exists a uniformly finite index set, $J_x$,
for each $x$ so that  
\begin{eqnarray}
 \lefteqn{\mbox{supp}(\sum_{i=1}^{\infty} \beta_i v_i) \cap \{y \in
 \R^m : |x-y| \leq 1\}}\nonumber\\
    &= & \mbox{supp}(\sum_{i\in J_x} \beta_i v_i) \cap \{y \in \R^m : |x-y|
  \leq 1\}. 
 \end{eqnarray}
For $\rho<m$ we have
\begin{eqnarray}
  \lefteqn{\int_{|x-y|\leq 1} \big|\sum_{i=1}^{\infty} \beta_i v_i(y)\big|^2\, |x-y|^{\rho -m} \,d^m y}\nonumber\\
   & = & \int_{|x-y|\leq 1} \big|\sum_{i\in J_x} \beta_i v_i(y)\big|^2\, |x-y|^{\rho -m}\, d^m y\nonumber\\
   & = & \int_{|x-y|\leq 1} \Big|\sum_{i\in J_x} \beta_i v_i(y)\; |x-y|^{\frac{\rho -m}{2}}\Big|^2 \,d^m y\nonumber\\
   & \leq & \Bigm(\sum_{i \in J_x} \Big(\int_{|x-y|\leq 1} \Big|\beta_i v_i(y) 
     |x-y|^{\frac{\rho -m}{2}}\Big|^2 \,d^m
   y\Big)^{\frac{1}{2}}\Bigm)^2 
 \end{eqnarray}
where in the last inequality the Minkowski or triangle inequality for
$L^2$ has been exploited. In a second step we get 
\begin{eqnarray}
  \lefteqn{\Bigm(\sum_{i \in J_x} \Big(\int_{|x-y|\leq 1} \Big|\beta_i v_i(y) 
     |x-y|^{\frac{\rho -m}{2}}\Big|^2 \,d^m
   y\Big)^{\frac{1}{2}}\Bigm)^2}\nonumber\\
    &\leq& (\sup_{i\in J_x} |\beta_i|)^2\, n_1^2\, \big(\max_{i\in J_x} M_{v_i,\rho}\big)^2
     \nonumber\\
&\leq& \|\beta\|_p^2\; n_1^2 \big(\max_{i\in J_x}
     M_{v_i,\rho}\big)^2\nonumber\\
&<&  \infty
 \end{eqnarray}
uniformly in $x$ as $\sup_{i\in J_x} |\beta_i| \leq \|\beta\|_\infty \leq \|\beta\|_p$.
Analogously one shows for $\rho\geq m$:
\begin{eqnarray}
  \int_{|x-y|\leq 1} \big|\sum_{i=1}^{\infty} \beta_i v_i(y)\big|^2 \,d^m y
   & = & \int_{|x-y|\leq 1} \big|\sum_{i\in J_x} \beta_i v_i(y)\big|^2\, d^m y\nonumber\\
   &\leq& \|\beta\|_p^2 \, n_1^2\, \big(\max_{i\in J_x} M_{v_i,\rho}\big)^2\nonumber\\
   &<&  \infty 
\end{eqnarray}
uniformly in $x$, which proves the statement.
\bewende

In the following we choose $H_0=-\Delta$. The Laplace operator is
selfadjoint on the Sobolev space $\W$ (see
e.\ g.\ \cite{RS2}). Furthermore, it can be inferred from slightly more
general results provided in \cite{Weid} that potentials from the
Stummel class with $\rho<4$ are defined on $\W$ and are
\tit{relatively bounded} with respect to $-\Delta$ with \tit{relative
  bound} zero.
\begin{satz}
 Let $\|\cdot\|,\|\cdot\|_2$ be the ($L^2$) Hilbert space and Sobolev
norm, respectively. For $\rho<4$ there exists a constant $C\geq0$ so
that\begin{equation} \|v\psi\| \leq C M_{v,\rho} \|\psi\|_2  \qquad
\forall\, v \in M_\rho(\R^m),\, \psi\in \W.\end{equation} Furthermore,
for all $\eta>0$ there exists a $C_{\eta}$ so that
\begin{equation}\|v\psi\| \leq \eta \|\psi\|_2 + C_{\eta} \|\psi\| 
\qquad \forall \psi\in \W.
\end{equation}
As the above Sobolev norm is equivalent to the {\em graph norm} of the
Laplacian it follows that $V$ is $-\Delta$-bounded with {\em relative
bound} zero.
\end{satz}

\begin{consequences} Under the above assumptions $H(\beta) = -\Delta
+ \sum_{i=1}^\infty \beta_i V_i$ is a closed respectively selfadjoint
operator on $\W$.
\end{consequences}

A slight extension then yields:
\begin{satz}
Under the above assumptions $H(\beta) = -\Delta + U +
\sum_{i=1}^\infty \beta_i V_i$ is a closed or selfadjoint operator on
$\W$ if $U$ is $-\Delta$-bounded with relative bound
zero. In the latter case $U$ and $V_i$ have to be
symmetric and $\beta_i$ have to be real. 
\end{satz}

So far the results on closedness or selfadjointness of Hamilton
operators and the corresponding classes of admissible potentials. In
the next two sections we are going to establish a
theory of \tit{analytic perturbation of spectra and operators} taking
place in \tit{infinitely many variables} at a time or variables varying in
a general Banach space upon this groundwork.

\section{Complex Analysis in Banach spaces}

In the second part of this paper we want to discuss analyticity
properties of eigenvalues of Hamilton operators $H(\beta)=H_0+
\sum_{i=1}^\infty \beta_i V_i$ with $H_0$ some unperturbed Hamiltonian
and $\beta_1, \beta_2, \ldots \in \C$. To do so one needs the notion
of infinite dimensional holomorphy.\\
Instead of dealing with infinitely many coupling parameters we will
frequently regard 
$\beta = (\beta_1, \beta_2, \ldots)$ as an element of a Banach space,
like e.g. $l^\infty$ and hence investigate perturbations in one
Banach space valued coupling parameter. While complex
analysis in one complex variable belongs to the standard repertoire
of the ordinary perturbation theory of operators we have to generalize
it in the way described above to complex analysis in Banach spaces.\\
As this is perhaps not so widely known we summarize definitions and theorems which will be
important in this enterprise. Many of the results can already be
found in \cite{Hille.Phillips}. As to more recent representations see
e.g. \cite{Mujica} or \cite{Zeidler}.

In what follows $X$ and $Y$ are infinite dimensional complex Banach
spaces and $U\subset X$ is an open set. One of the difficulties of complex
analysis in Banach spaces is a suitable
definition of power series and differentiability.

\begin{defi}
 A {\em formal power series} from $X$ to $Y$ at $a\in X$ is a series of
 symmetric, $m$-linear mappings $A_m:X^m \to Y$ of the form
 \begin{equation} \sum_{m=0}^\infty A_m(x-a)^m\end{equation}
 with
 $A_m(x-a)^m:=A_m(\underbrace{x-a,\ldots,x-a}_{m-\mbox{\footnotesize times}})$.
\end{defi}

The radius of convergence of the power series $\sum_{m=0}^\infty
A_m(x-a)^m$ is the supremum of all $r\geq 0$ so that the series
converges uniformly in the closed ball $\overline{B}(a,r)$.
In analogy to the formula of Cauchy-Hadamard the radius of convergence
$R$ is given by
\begin{equation} \frac{1}{R} = \limsup_{m\to \infty} \|A_m\|^{\frac{1}{m}}\end{equation}
with $\|A_m\| := \sup_{\|x_1\|=\ldots=\|x_m\|=1}
\|A_m(x_1,\ldots,x_m)\|$ and $\frac{1}{0} := \infty$ as well as
$\frac{1}{\infty} := 0$. The series converges absolutely and
uniformly in $\overline{B}(a,r)$ if $0\leq r<R$.\\

With this notion of power series it is possible to introduce analytic
mappings in Banach spaces.

\begin{defi} \label{Definition 5.3}
  A map $f: U\to Y$ is called {\em analytic} if for each $a\in U$
  exists a ball $B(a,r) \subset U$ and a sequence of symmetric,
  $m$-linear, continuous mappings $A_m:X^m \to Y$ so that
 \begin{equation}f(x)=\sum_{m=0}^\infty A_m(x-a)^m\end{equation}
  for all $x\in B(a,r)$.
\end{defi}

The sequence of mappings $A_m$ is uniquely determined by $f$ and $a$. We
will frequently suppress the explicit dependence on $f$ or $a$ and set
\begin{equation} A^m f(a):=A_m.\end{equation}
\begin{eqnarray} \label{Formel 5.B}
 f(x)=\sum_{m=0}^\infty A^m f(a)(x-a)^m
\end{eqnarray}
is called {\em Taylor series} of $f$ at $a$.

Many important theorems in complex analysis in Banach spaces can
be shown to hold by reducing the problems to well-known results of
complex analysis in one or several complex variables.
\begin{defi} \label{G-analytisch}
  A map $f:U\to Y$ is called {\em G-analytic} if the mapping $\lambda \mapsto
    f(a+\lambda b)$ is analytic for all $a\in U$ and $b\in X$ on the
    open set $\{\lambda\in \C : a+ \lambda b \in U\}$.
 It is called {\em weakly analytic} if $g\circ f$ is analytic for all
    $g\in Y^\prime$, the dual space of $Y$.
\end{defi}

The following generalized Cauchy integral formula is useful.

\begin{satz}  \label{Cauchysche Integralformel}
  Let $f:U\to Y$ be analytic, $a\in U$, $t\in X$ and $r>0$ so that
  $a+ \zeta t\in U$ for all $\zeta \in \overline{{\cal
      U}(0,r)}\subset \C$. Then for all $\lambda\in {\cal
    U}(0,r)\subset \C$ the {\em Cauchy integral formula}
 \begin{equation}f(a+ \lambda t)= \frac{1}{2\pi i} \int_{|\zeta| = r} 
           \frac{f(a+\zeta t)}{\zeta-\lambda}\, d\zeta\end{equation}
 holds. Further
 \begin{equation}\frac{d}{d\lambda} f(a+\lambda t) = \frac{1}{2\pi i}\int_{|\zeta| = r} 
           \frac{f(a+\zeta t)}{(\zeta-\lambda)^2}\, d\zeta\end{equation}
 is valid.
\end{satz}

The integration paths are always positively oriented. An analogous
formula exists for higher derivatives. 

With the help of the generalized Cauchy integral formula the
following relations between the different notions of
differentiability can be shown:

\begin{satz} \label{rem}
\begin{eqnarray*}
f\ \mbox{is analytic.}& \Leftrightarrow & f\ \mbox{is continuous and G-analytic.}\\
& \Leftrightarrow & f\ \mbox{is locally bounded and G-analytic.}\\
& \Leftrightarrow  & f\ \mbox{is weakly analytic.}
\end{eqnarray*}
\end{satz}

After discussing the notion of analytic or holomorphic functions in
the sense of
power series, i.e. the point of view adopted by Weierstrass, we now turn to
the notion of complex differentiability, i.\ e. the Riemannian point
of view.  

\begin{defi} \label{Definition 5.6}
  A map $f:U\to Y$ is called {\em differentiable}
  (Fr\'echet-differentiable, complex differentiable or differentiable
  in norm) if for all $a\in U$ there exists a mapping $A\in {\cal L}(X,Y)$
  so that
 \begin{equation}\lim_{h\to 0} \frac{\|f(a+h) -f(a) - A(h)\|}{\|h\|} =0.\end{equation}
\end{defi}

Alternatively a map $f:U\to Y$ is called differentiable if for all
$a\in U$ a mapping $A\in {\cal L}(X,Y)$ exists so that 
\begin{equation} f(a+h) - f(a) = A(h) + o(\|h\|)\end{equation}
for all $h$ in a neighborhood of zero. Here $r(h)=o(\|h\|)$ is an
abbreviation for a mapping $r:{\cal U}(0) \subseteq X \to Y$ with
$\frac{r(h)}{\|h\|} \to 0$ if $h\to 0$.

As a side remark we want to mention some further results.

\begin{enumerate}
 \item Every differentiable map $f:U\to Y$ is continuous.
 \item The mapping $g\mapsto g^{-1}$ is differentiable for every
   invertible map $g\in {\cal L}(Y)$. We will use this in connection with
   resolvents in section 5.
 \item The mapping $A\in {\cal L}(X,Y)$ of Definition
    \ref{Definition 5.6} is uniquely determined by $f$ and $a$. It is
    called {\em derivative} of $f$ in $a$ and is often written in the form
    \begin{equation}Df(a):= A.\end{equation}
\end{enumerate}

Every differentiable map $f:U\to Y$ induces a mapping $Df:U\to {\cal
  L}(X,Y)$. 
As in finite dimensional Banach spaces sum rule, product rule and
chain rule are valid.
In the proofs of section 5, where we generalize perturbation
theory to coupling parameters in Banach spaces, we often use the
equivalence between analyticity and complex differentiability.\\
Analogous to complex analysis in one complex variable it is shown that
analyticity also implies that the map is infinitely often complex
differentiable. Here higher derivatives are defined recursively,
i.\ e.\ $f:U\to Y$ is {\em $k$-times differentiable} if $f$ is a
$(k-1)$-times differentiable mapping and if the $(k-1)$st derivative
$D^{k-1} f: U \to {\cal L}(X^{k-1},Y)$  is differentiable. A map is
called {\em infinitely often differentiable} if it is $k$-times
differentiable for each $k\in \N$ (we set $D^0 f=f$).

One can show that every $m$-linear mapping 
$D^m f(a) \in {\cal L}(X^m,Y)$ is symmetric for all $a\in U$. With
this it is possible to prove the following important theorem.

\begin{satz} \label{Satz 5.12}
 For a map $f:U\to Y$ the following statements are equivalent:
 \begin{enumerate}
  \item $f$ is analytic.
  \item $f$ is complex differentiable.
  \item $f$ is infinitely many times complex differentiable.
 \end{enumerate}
If one of these conditions is fulfilled, 
 \begin{equation} D^m f(a) = m!\,A^m f(a)\end{equation}
holds.
\end{satz}

With this we get the Taylor series of $f$ in $a$ 
\begin{equation} f(x)= \sum_{m=0}^\infty \frac{1}{m!} D^m f(a) (x-a)^m\end{equation}
for all $x\in B(a,r)\subset U$ with a certain $r\geq 0$.\\

In section 5 we will use the notion of a differentiable map, which depends
on a variable in a Cartesian product of Banach spaces. We want to
show that the resolvent of the Hamilton operator $H(\beta)$,
\begin{equation}\big(H(\beta)-\lambda\big)^{-1},\end{equation}
is jointly differentiable in $(\beta,\lambda)$ with $\lambda$ being an
element of the resolvent set.\\
The Cartesian product $X\times Y$ of two Banach spaces $X$ and $Y$
becomes a Banach space by component-wise addition and scalar
multiplication and  with the norm $ \|(x,y)\| := \|x\| +\|y\| $ for $
(x,y) \in X\times Y$. Let $U\subset X\times Y$ be open and $Z$ another
complex Banach space. Then $f:U\to Z$ is called differentiable in
analogy to Definition \ref{Definition 5.6} if for all $(a,b) \in U$
there exists a mapping $A\in {\cal L}(X\times Y, Z)$ so that 
\begin{equation}\lim_{(h,k)\to (0,0)} \frac{\|f(a+h,b+k) -f(a,b) - A(h,k)\|}{\|(h,k)\|} =0\end{equation}
or equivalently
\begin{equation}f(a+h,b+k) = f(a,b) + A(h,k) + o(\|(h,k)\|)\end{equation}
holds.
The mapping $A$ is called derivative of $f$ in $(a,b)\in U$ and is
written like $Df(a,b):=A$.\\

To prove the differentiability of a mapping in e.g. two variables,
one can introduce partial derivatives as in finite dimensional spaces.

\begin{satz} \label{Satz 5.13}
 A map $f:U\subset X\times Y \to Z$ is differentiable if
 $f_1(x):=f(x,y)$ and $f_2(y):=f(x,y) $ are differentiable and if the  
 derivatives $D_1 f(a,b) := D f_1(a)$ and $D_2 f(a,b) := D f_2(b)$ are
 continuous in $(a,b)$. Then for each $(a,b)\in U$
 \begin{equation}D f(a,b)(h,k) = D_1 f(a,b) h + D_2 f(a,b) k\end{equation}
 holds for all $(h,k)\in X\times Y$.
\end{satz}

\section{Analytic perturbation theory in coupling parameters in Banach
spaces}

As the model Hamiltonian of section 1 suggests, we have to focus our
attention on the perturbation theory of operators in infinite many
complex coupling parameters. In the following we consider the
$\beta_i$'s as an element of a sequence space $\lp$. Of particular
interest is the space $l^{\infty}$, i.e. the sequences which are
uniformly bounded.\\
As we have already remarked in section 4, we treat the
sequence-space-valued coupling parameters in a more abstract way by
regarding them as coupling parameters in a general complex Banach space.
\\
In the first subsection we define analytic families and prove a generalization of a theorem of Kato and Rellich about the
behavior of isolated, non-degenerate eigenvalues and their
eigenfunctions.\\
The second subsection deals with other notions of analytic
families. We investigate in particular analytic families of type (A)
and explore their 
relation to analytic families of the first subsection.\\
In the third subsection we show that relatively bounded
perturbations are analytic families in our generalized sense. 
This then enables us to apply the machinery developed above to the
model Hamiltonian of section 1.

\subsection{Generalization of a theorem of  Kato and Rellich}

One of the goals of analytic perturbation theory is the representation of
eigenvalues and eigenfunctions as power series in the complex
coupling parameter. Therefore the functions under discussion have to be
analytic in the coupling parameter. One hopes that the
eigenvalues and eigenfunctions are analytic if the corresponding
Hamilton operator depends analytically on the coupling parameter in a certain
way.\\
As we have explained in section 4 it suggests itself to investigate
analytic mappings between a Banach space of coupling constants and a
Banach space of operators, for example the bounded operators.\\
For unbounded (Hamilton) operators the situation is slightly different as the
set of unbounded operators is not automatically a Banach space. It is
however possible to
metrize the set of closed operators and to define analytic families
via a generalized convergence \cite[p.\ 197]{Ka}. In this paper we
use an equivalent definition according to
\cite[p.\ 14]{Reed.Simon.4}. 
In this approach analyticity of the
corresponding resolvents is demanded, such that the problem is reduced to
the case of bounded operators.\\
In the following `$\beta$ near $\beta_0$' always means that $\beta$
is an element of a suitable neighborhood of $\beta_0$. If not stated
otherwise, the operators $T(\beta)$ are defined on a Banach space 
$Y$. $X$ is always assumed to be a complex Banach space and $U\subset
X$ to be open and connected.

\begin{defi} \label{3.3}
  An operator-valued mapping $T(\cdot)$ on $U$ is called an {\em analytic
   family} or an {\em analytic family in the sense of Kato} if and only if
 \begin{enumerate}
  \item For each $\beta \in U$, the operator $T(\beta)$ is closed and  
    has a non-empty resolvent set, i.e. $\rho (T(\beta)) \neq
    \emptyset$.
  \item For every $\beta_0 \in U$ a $\lambda_0 \in \rho
    (T(\beta_0))$ exists so that $\lambda_0 \in \rho (T(\beta)) $ if $
    \beta$ is near $\beta_0$ and so that the resolvent
    $(T(\beta) - \lambda_0)^{-1} $ is an analytic operator-valued
    mapping of $\beta$ in a neighborhood of $\beta_0$.
 \end{enumerate}
\end{defi}

The following investigations show that this definition is convenient
and allows to derive results about the behavior of eigenvalues and
eigenfunctions, like e.g. the generalized theorem of Kato and Rellich.\\ 
For this we need the analyticity of the resolvent in both variables
$(\beta,\lambda)$.

\begin{lemma} \label{Satz 6.3}
 If $T(\cdot)$ is an analytic family on $U,$
 \begin{equation} \Gamma := \{ (\beta, \lambda) : \beta \in U,\, \lambda \in 
                                                    \rho (T(\beta))\} \end{equation} 
 is open in $X \times \C$. The resolvent $ (T(\beta) - \lambda)^{-1}$,
 which is defined on $ \Gamma$, is analytic in $(\beta, \lambda)$.
\end{lemma}

The proof, which is inspired by \cite[p.\ 14]{Reed.Simon.4}, can be
found in \cite{Schloem}. It exploits the equivalence between analyticity
and complex differentiability. The resolvent is analytic in $(\beta,
\lambda) \in \Gamma$ if it is differentiable in each variable
and if the partial derivatives are continuous in $(\beta,\lambda)$
(Theorem \ref{Satz 5.13}). In particular we use the differentiability of the mapping
$g\mapsto g^{-1}$.

With the help of this it is possible to generalize the theorem of Kato and Rellich
of perturbation theory in one complex parameter
\cite[p.\ 15]{Reed.Simon.4} to coupling parameters in a complex Banach
space. 

\begin{satz} \label{Kato-Rellich}
 Let $T(\cdot)$ be an analytic family
 in $\beta\in U$. Suppose that $E_0$ is an
 isolated, non-degenerate eigenvalue of $T(\beta_0)$, then the
 following is valid: 
 \begin{enumerate}
  \item For $\beta$ near $\beta_0$, there is exactly one isolated,
    non-degenerate point $E(\beta)$ in $\sigma(T(\beta))$
    near $E_0$. $E(\beta)$ is an analytic map of $\beta$ for $\beta$
    near $\beta_0$. 
  \item There is an analytic eigenvector $\psi(\beta)$ of
    $T(\beta)$ for $\beta$ near $\beta_0$.
 \end{enumerate}
\end{satz}

\bew
 For $E_0$ being a discrete eigenvalue of $T(\beta_0)$, one can find
 $r >0$ so that $\{\lambda\in \C : |\lambda-E_0| \leq r\} \cap
 \sigma(T(\beta_0)) = \{E_0\}$. 
 The circle $\{\lambda\in \C : |\lambda-E_0| = r\}$ is compact in $\C$ and 
 a subset of $\rho(T(\beta_0))$. According to Lemma \ref{Satz 6.3} the
 set $\Gamma = \{(\beta,\lambda) : \beta \in U \subset X,\, \lambda\in
 \rho(T(\beta))\}$ is open in $X\times \C$. Therefore $\delta >0$
 exists so that $\lambda \in \rho(T(\beta))$ if $|\lambda-E_0| = r $ and 
 $\|\beta - \beta_0\| \leq \delta$. Then the resolvent
 $(T(\beta)-\lambda)^{-1}$ is analytic in $(\beta,\lambda)$. Let
 $W:=\{\beta \in X : \|\beta-\beta_0\| \leq \delta\}.$ Then 
 \begin{equation}P(\beta)= -(2\pi i)^{-1} \int_{|\lambda-E_0| = r}
                              \big(T(\beta)-\lambda \big)^{-1} \, d\lambda\end{equation}
 exists for all $\beta \in W$ and is analytic in $\beta$ if $\beta\in 
 W\subset X$.\\
 Since the eigenvalue $E_0$ of $T(\beta_0)$ is non-degenerate, the
 corresponding projector is one-dimensional. Using a lemma 
 in \cite[p.\ 14]{Reed.Simon.4} we know that all projectors
 $P(\beta)$  are one-dimensional if $\beta \in W$. According to
 theorem XII.6 in \cite[p.\ 13]{Reed.Simon.4}, which is also valid for
 operators in Banach spaces, there is exactly one non-degenerate
 eigenvalue $E(\beta)$ of $T(\beta)$ with $|E(\beta)-E_0| < r$ if
 $\beta \in W$.\\  
 Let $\psi_0$ be the corresponding eigenvector of $E_0$. Then $P(\beta)\psi_0
 \neq 0$ if $\beta$ is near $\beta_0$ because $P(\beta)\psi_0 \rightarrow
 \psi_0 $ for $\beta \rightarrow \beta_0$. As $P(\beta)\psi_0$ is an
 eigenvector of the operator $T(\beta)$ in $Y,$ we have for all $\varphi\in
 Y^\prime,$ the dual space of $Y,$ 
 \begin{eqnarray}
  \varphi (P(\beta)\psi_0)
    &=& \varphi \Big(\big(T(\beta) - E_0 -r\big)^{-1} \big(T(\beta)-E_0 -r\big)
                      P(\beta)\psi_0\Big)\nonumber\\
    &=& (E(\beta) - E_0 -r) \; \varphi \Big(\big(T(\beta) - E_0 -r\big)^{-1}
                      P(\beta)\psi_0\Big).
 \end{eqnarray}
 Hence
 \begin{equation}(E(\beta) - E_0 -r)^{-1}=\frac{ \varphi \Big(\big(T(\beta) - E_0 -r\big)^{-1}
                      P(\beta)\psi_0\Big)}{\varphi \big(P(\beta)\psi_0\big)}\end{equation}
 and $(E(\beta) - E_0 -r)^{-1}$ is weakly analytic if
 $\beta\in W$ and therefore analytic as explained in section 4. Because
 the mapping $g\mapsto g^{-1}$ is analytic, $E(\beta)$ is analytic in
 $\beta \in W$.\\ 
 Define $\psi(\beta):=P(\beta)\psi_0$, then $\psi(\beta)$ is an analytic
 eigenvector of $T(\beta)$ if $\beta \in W$. 
\mbox{}\bewende

\subsection{Analytic families of type (A)}

As the last theorem shows the notion of analytic families in the sense
of Kato is also convenient for coupling parameters in general Banach
spaces. As it is frequently difficult to varify directly that a
given family of operators is an {\em analytic} family, other notions
of analytic families are introduced.

In this paper we define analytic families of type (A) for operators
depending on a parameter varying in a Banach space. It is then
possible to show that analytic families of this type are analytic
families in the more general sense of Kato. This is useful because it
is usually easier to prove that a family of operators is analytic of
type (A).

\begin{defi} \label{Typ A}
 For each $\beta \in U$, let $T(\beta) : {\cal D}(T(\beta))
 \subset Y \to Y$ be a closed operator with non-empty resolvent
 set. $T(\cdot)$  is called an {\em analytic family of type (A)} if and
 only if
 \begin{enumerate}
  \item The domain ${\cal D}:={\cal D}(T(\beta))$ does not depend on
    $\beta \in U$.
  \item $T(\beta)\psi$ is an analytic map in $\beta \in U$ for all
    $\psi \in {\cal D}$. 
 \end{enumerate}
\end{defi}

In order to infer the more general property from this we prove the
analyticity of the resolvent with the help of the `strong'
analyticity of the operators. In a first step we construct bounded operators
from the closed operators and use the following lemma.

\begin{lemma} \label{Lemma 6.8}
 Let $X$, $Y$ and $Z$ be complex Banach spaces and let $U\subset X$ be open.
 If $\tilde{T}(\beta) \in {\cal L}(Z,Y)$ and if $\tilde{T}(\beta)\psi$ 
 is analytic in $\beta\in U$ for all $\psi \in Z$, then
 $\tilde{T}(\beta)$ is analytic in $\beta \in U$.
\end{lemma}

\bew
 Let $\beta\in U$, $t\in X$ and $M= \{\zeta \in \C : \beta + 
 \zeta t \in U\}$. Let $\Gamma \subset M$ be a mathematically positive oriented
 circle in $\zeta$. The Cauchy integral formula (Theorem \ref{Cauchysche 
 Integralformel}) yields
 \begin{eqnarray}
  \lefteqn{\frac{1}{h} \left(\tilde{T}(\beta + (\zeta + h) t)\psi 
          - \tilde{T}(\beta + \zeta t)\psi\right)
          - \frac{d}{d\zeta} \tilde{T}(\beta + \zeta t)\psi}\nonumber\\
  &=& \frac{1}{2\pi i} 
       \int_\Gamma \left(\frac{1}{h} \left( \frac{1}{\zeta^\prime -(\zeta + h)}
                                   - \frac{1}{\zeta^\prime - \zeta}\right)    
                     - \frac{1}{(\zeta^\prime - \zeta)^2}\right) 
                   \tilde{T}(\beta + \zeta^\prime t) \psi \, d\zeta^\prime\nonumber\\ 
  &=& \frac{1}{2 \pi i} 
        \int_\Gamma \frac{h}{\left(\zeta^\prime - (\zeta + h)\right)(\zeta^\prime - \zeta)^2}
                   \,\tilde{T}(\beta + \zeta^\prime t) \psi \, d\zeta^\prime.
 \end{eqnarray}

 As all analytic functions are G-analytic,  $\tilde{T}(\beta)\psi$ is
 G-analytic (Theorem \ref{rem}). Hence $\tilde{T}(\beta + \zeta^\prime t)
 \psi$ is continuous  in $\zeta^\prime\in M$. Since $\Gamma$ is
 compact, for each $\psi \in Z$ a number $C_\psi$ exists so that
 \begin{equation} \|\tilde{T}(\beta + \zeta^\prime t)\psi\| \leq C_\psi\end{equation}
 for all $\zeta^\prime \in \Gamma$. According to the uniform
 boundedness theorem a $C\in \R$ exists so that
 \begin{equation} \sup_{\zeta^\prime \in \Gamma} \|\tilde{T}(\beta + \zeta^\prime t)\|
            \leq C.\end{equation}
 Therefore one gets
 \begin{eqnarray}
 \lefteqn{\|\frac{1}{h} \left(\tilde{T}(\beta + (\zeta + h) t)\psi 
           - \tilde{T}(\beta + \zeta t)\psi\right)
           - \frac{d}{d\zeta} \tilde{T}(\beta + \zeta t)\psi\|}\nonumber\\
  &\leq& \frac{1}{2 \pi} C \|\psi\| \int_\Gamma 
             \left|\frac{h}{(\zeta^\prime - (\zeta + h))
                                (\zeta^\prime - \zeta)^2}\right|
              \, d\zeta^\prime,
 \end{eqnarray}
 hence the estimate
 \begin{eqnarray}
 \lefteqn{\|\frac{1}{h} \left(\tilde{T}(\beta + (\zeta + h) t) 
          - \tilde{T}(\beta + \zeta t)\right)
          - \frac{d}{d\zeta} \tilde{T}(\beta + \zeta t)\|}\nonumber\\
    &\leq& \frac{1}{2 \pi} C \int_\Gamma
              \left|\frac{h}{(\zeta^\prime - (\zeta + h))
                                    (\zeta^\prime - \zeta)^2}\right|
              \, d\zeta^\prime
 \end{eqnarray}
 holds.
 The integral vanishes in the limit $h\to 0$. Therefore
 $\tilde{T}(\beta)$ is G-analytic in $\beta \in U$. $\tilde{T}(\beta)$
 is analytic if, in addition, $\tilde{T}(\beta)$ is locally bounded
 (Theorem \ref{rem}). The local boundedness follows from the
 uniform boundedness principle by means of the continuity of the mapping
 $\tilde{T}(\beta)\psi$. For each $\psi \in Z$ and every compact set 
 $\Gamma \subset U$ a $c_\psi$ exists so that
 $\|\tilde{T}(\beta)\psi\| \leq c_\psi$ is valid for all $\beta \in
 \Gamma$. Hence $c\in \R$ exists with
 $\sup_{\beta\in \Gamma} \|\tilde{T}(\beta)\| \leq c$.
\bewende

This yields the following important theorem (cf. \cite[p.\ 375]{Ka}).

\begin{satz} \label{Satz 6.9}
 Every analytic family of type (A) is an analytic family in the sense
 of Kato.
\end{satz}

\bew
 Let $\beta_0 \in U\subset X$. Let
 $T(\beta): {\cal D}\subset Y \to Y$  be an analytic family of type (A)
 in $\beta \in U$. Hence $T(\beta_0)$ is a closed operator with
 $\rho(T(\beta_0))\neq  \emptyset$. By introduction of the graph norm
 \begin{equation} |\!|\!| \psi |\!|\!| = \|\psi\| +
   \|T(\beta_0)\psi\|\end{equation} the domain ${\cal D}$ of this
 operator is converted into a Banach space 
 \begin{equation}\tilde{D}:=({\cal D},|\!|\!| \cdot |\!|\!|).\end{equation}
 Let $\iota$ be the embedding operator from $\tilde{D}$
 in $Y$. $\iota$ is bounded because $\|\iota \psi\|= \|\psi\|
 \leq |\!|\!| \psi |\!|\!| $ holds. \\
  We now consider the operator $T(\beta)$ from $\tilde{D}$ to $Y$ and
  call this operator $\tilde{T}(\beta)$.
 \begin{eqnarray}
  \tilde{T}(\beta): \tilde{D}&\to& Y,\\
                  \psi &\mapsto& \tilde{T}(\beta)\psi=T(\beta)\,\iota\,\psi.
 \end{eqnarray} 
 $\tilde{T}(\beta)$ is a closed operator because $T(\beta)$ is closed
 and $\iota$ is continuous. $\tilde{T}(\beta)$ is defined on the whole
 $\tilde{D}$. Therefore $\tilde{T}(\beta)$ is bounded according to the
 closed graph theorem, i.e.  $\tilde{T}(\beta) \in {\cal L}(\tilde{D},Y)$.\\
 $\tilde{T}(\beta)\psi= T(\beta)\,\iota\,\psi =
 T(\beta)\psi$ is analytic in $\beta \in U$ for all $\psi \in
 \tilde{D}$. Therefore $\tilde{T}(\beta)$ is analytic in $\beta \in U$
 (Lemma \ref{Lemma 6.8}).\\
 Let $\lambda_0\in \rho(T(\beta_0))$. It has to be shown that
 $\lambda_0 $ is an element of $\rho(T(\beta))$ and that
 $(T(\beta)-\lambda_0)^{-1}$ is an analytic map in $\beta$ for $\beta$
 near $\beta_0$.
 The map $T(\beta_0) - \lambda_0 : {\cal D} \to Y$ is bijective because
 $\lambda_0\in  \rho(T(\beta_0))$. The same holds for $\iota:
 \tilde{D}\to {\cal D}$. Therefore $\big(T(\beta_0) 
 -\lambda_0\big)\iota = \tilde{T}(\beta_0) - \lambda_0 \,\iota$
 is invertible and 
 \begin{equation}\big(\tilde{T}(\beta_0)-\lambda_0\, \iota\big)^{-1} \in {\cal
   L}(Y,\tilde{D}).\end{equation}
 
 As the set of invertible, continuous and linear operators on $Y$
 is open (see e.\ g.\ \cite[p.\ 9]{Takesaki}),
 \begin{equation} \big(\tilde{T}(\beta) - \lambda_0\, \iota\big)^{-1} \in {\cal
   L}(Y,\tilde{D})\end{equation}
 and $(\tilde{T}(\beta) - \lambda_0\, \iota)^{-1}$ is
 analytic in $\beta$ for $\beta$ near $\beta_0$. Therefore    
 \begin{equation} \big(T(\beta)-\lambda_0\big)^{-1} = \iota\, \big(\tilde{T}(\beta)
 - \lambda_0 \,\iota\big)^{-1} \end{equation} 
 is bounded and analytic in $\beta$ (note that we modified the
 standard textbook proof which does not seem to be directly applicable
 to our more general situation).
\bewende

The inversion of this theorem is not valid as already has been shown
by a counterexample for
complex coupling parameters \cite[p.\ 376]{Ka} or
\cite[p.\ 20]{Reed.Simon.4}. An analytic family in the sense of Kato
can have a domain which depends on $\beta$.

\subsection{Perturbation theory of Hamilton operators in infinitely many
  complex coupling parameters}

In quantum mechanics the (Hamilton) operators are typically of the form
\begin{equation}H(\beta)=H_0 + V(\beta).\end{equation}

\begin{satz} \label{Satz 6.10}
 Let $X$ and $Y$ be complex Banach spaces;
 let $U\subset X$ be open and connected. Suppose that $H_0$ is a
 closed operator on 
 ${\cal D}(H_0) \subset Y$ and that for each $\beta\in U$, $V(\beta)$ is
 relatively $H_0$-bounded  with $H_0$-bound smaller than one. Furthermore
 let $V(\beta)\psi$ be analytic in $\beta\in U$ for all $\psi\in {\cal
   D}(H_0)$  and let the resolvent set $\rho(H(\beta))$ of
  \begin{equation}H(\beta)= H_0 + V(\beta), \quad \beta\in U,\end{equation}
 be non-empty. Then $H(\cdot)$ is an analytic family.
\end{satz}

\bew
 It is sufficient to prove that $H(\cdot)$ is an analytic family of
 type (A) (Theorem \ref{Satz 6.9}).
 According to a well-known stability theorem \cite[p.\ 190]{Ka},
 $H(\beta)$  is a closed operator for all $\beta \in U$.
 The domain ${\cal D}(H(\beta))= {\cal D}(H_0)$ does not depend on
 $\beta$. Because $V(\beta)\psi$ is an analytic map,
 $H(\beta)\psi$ is also analytic in $\beta\in U$ for all $\psi \in
 {\cal D}(H_0)$. Therefore all conditions of an analytic family of
 type (A) are fulfilled.
\mbox{}\bewende

\begin{bem} \label{montag}
 Let $H_0$ be a selfadjoint operator on $ {\cal D}(H_0)\subset
 {\cal H}$ and let $V$ be relatively $H_0$-bounded with $H_0$-bound
 zero. Then the resolvent set $\rho(H_0+ V)$ is not empty.
\end{bem}

\bew
 By definition we have
 \begin{eqnarray} \label{formelpaper}
 \|V\psi\| \leq a\|\psi\| + b \|H_0\psi\| \quad \forall \psi \in {\cal
   D}(H_0).
 \end{eqnarray}
 As the $H_0$-bound of $V$ is zero, $b$ can be chosen arbitrarily
 small. The spectrum of the selfadjoint operator $H_0$ is
 real. Therefore $\lambda \in \rho(H_0)$ with a sufficiently large
 imaginary part exists so that
 \begin{equation}a \sup_{E\in \sigma(H_0)} |E-\lambda|^{-1} +
       b \sup_{E\in \sigma(H_0)} |E| |E-\lambda|^{-1} <1\end{equation}
 holds. According to \cite[p.\ 214 and 272]{Ka} $\lambda$ is an
 element of $\rho(H_0 + V)$.
\bewende

In the first part of the paper we investigated Hamilton operators of
the form
\begin{equation}H(\beta)=H_0 + \sum_{i=1}^\infty \beta_i V_i.\end{equation}
$V(\beta)=\sum_{i=1}^\infty \beta_i V_i$ is analytic in
$\beta=(\beta_1,\beta_2,\ldots) \in l^\infty(\C)$  for all $\psi \in {\cal
  D}(H_0)$ because $\sum_{i=1}^\infty \beta_i V_i$ is
continuous and linear in $\beta$ for all $\psi \in {\cal
  D}(H_0)$. Bounded operators are relatively $H_0$-bounded with $H_0$-bound zero. Therefore we get the following corollary.

\begin{koro} \label{nachts}
  Suppose $H_0$ to be a selfadjoint operator with ${\cal D}(H_0) \subset
  L^2(\R^m)$. Let $V_i$ be symmetric, bounded operators in $L^2(\R^m)$ with
  $\|V_i\| \leq v$ for all $i\in\N$, which fulfill the finite
  intersection property. Let $\beta\in l^\infty(\C)$. 
  Then
  \begin{eqnarray}
  H(\beta)= H_0 + \sum_{i=1}^\infty \beta_i V_i
  \end{eqnarray}
  is an analytic family.
\end{koro}

For Hamilton operators with an infinite sum of Stummel-class
potentials we get a corresponding result.

\begin{koro} \label{borch}
 Let $\rho<4$ and $\beta\in\lp$. Let $\{V_i\}_{i\in\N}$
 be multiplication operators in $L^2(\R^m)$ so that the finite
 intersection property is fulfilled and so that $v_i\in\m$ with
 $\sup_{i\in\N} M_{v_i,\rho} < \infty$. 
 Then
 \begin{eqnarray}
  H(\beta) = -\Delta + \sum_{i=1}^\infty \beta_i V_i
 \end{eqnarray}
 is an analytic family.
 If in addition $U$ is a symmetric, $-\Delta$-bounded
 operator with $-\Delta$-bound zero, the Hamilton operators
 $H(\beta)=-\Delta + U + \sum_{i=1}^\infty \beta_i V_i$ are an
 analytic family.
\end{koro}

\bew
 Since $-\Delta$ is a selfadjoint operator on $\W$, $-\Delta + U$
 is selfadjoint on $\W$. In section 3 we proved that 
 $V(\beta):=\sum_{i=1}^\infty \beta_i V_i$ is an element of the
 Stummel-class $\m$. Therefore $V(\beta)$ is relatively bounded with
 respect to $-\Delta + U$  with $(-\Delta + U)$-bound zero.
\bewende

\end{document}